# Triple equivalence for the emergence of biological intelligence

Takuya Isomura

Brain Intelligence Theory Unit, RIKEN Center for Brain Science, 2-1 Hirosawa, Wako, Saitama 351-0198, Japan

Corresponding author email: takuya.isomura@riken.jp

**Abstract**

Characterising the intelligence of biological organisms is challenging. This work considers intelligent algorithms developed evolutionarily within neural systems. Mathematical analyses unveil a natural equivalence between canonical neural networks, variational Bayesian inference under a class of partially observable Markov decision processes, and differentiable Turing machines, by showing that they minimise the shared Helmholtz energy. Consequently, canonical neural networks can biologically plausibly equip Turing machines and conduct variational Bayesian inferences of external Turing machines in the environment. Applying Helmholtz energy minimisation at the species level facilitates deriving active Bayesian model selection inherent in natural selection, resulting in the emergence of adaptive algorithms. In particular, canonical neural networks with two mental actions can separately memorise transition mappings of multiple external Turing machines to form a universal machine. These propositions were corroborated by numerical simulations of algorithm implementation and neural network evolution. These notions offer a universal characterisation of biological intelligence emerging from evolution in terms of Bayesian model selection and belief updating.

**INTRODUCTION**

Characterising the intelligence of biological organisms is challenging yet crucial. Although a conclusive definition of intelligence remains to be established, various intelligent functions are computable and can be implemented as algorithms [1]. This might be obvious considering that the dynamics of the fundamental units of the brain—i.e., neurons [2–4] and synapses [5–7]—can be expressed as algorithms. Moreover, biological intelligence has been shaped through evolution [8], driven by the preferential survival of organisms who successfully reproduce and contribute more offspring (i.e., genes) to subsequent generations. These features comprise requisites for biologically plausible algorithms. However, a precise characterisation of the biological intelligence that stems from evolution is yet to be established. This work addresses the aforementioned issue by characterising algorithms formed through natural selection in terms of variational Bayesian inference of external milieu states.

The brain can be characterised in terms of dynamical systems of neural circuits [9–11], agents that perform Bayesian inferences [12,13], and machines that perform computations. Thus, constructing a theory of brain intelligence requires the integration of these perspectives. The first two are closely related to optimisations and widely expressed as the minimisation of cost or energy functions, wherein a gradient descent on cost functions furnishes biologically plausible algorithms [14–16]. The free-energy principle has been proposed to account for the perception, learning, and actions of biological organisms in terms of variational Bayesian inference and free energy minimisation [17,18]. Although its physiological bases have been debated, recent works have shown that any neural network that minimises its cost function can be cast as performing variational Bayesian inference [19–21]. This is attributed to the natural equivalence between the Helmholtz energy implicit in neural networks and variational free energy under a class of generative models [21].

Previous works have established that the dynamics of canonical neural networks are formally homologous to Bayesian belief updates under a specific yet sufficiently generic class of partially observable Markov decision processes (POMDPs). These networks have been shown to perform causal inference [19], rule learning [22], and planning [20]. Moreover, the consistency between such model dynamics and experimental observations has been demonstrated using in vitro neural networks [23–25], wherein canonical neural networks can quantitatively predict the self-organisation of in vitro neural networks when assimilating sensory data, validating their biological plausibility [25]. These works licence the adoption of the free-energy principle as a universal characterisation of neural networks.

Having said this, previous computational models in neuroscience have addressed limited algorithms, which rely on specific (generative model) architectures assigned a priori and thus have limited flexibility. In contrast, any algorithm can be implemented in the form of a Turing machine [1], which is a simple mathematical model representing an automatic machine that can perform any computation comprising combinations of symbolic operations. Previous works have developed artificial neural networks combined with external memory [26,27] and spiking neural networks with a specific architecture [28–30] to implement Turing machines. However, neuronal substrates of external memory—available for reading and writing—remain unelucidated. Moreover, how biological agents can evolutionarily acquire (i.e., self-organise) these architectures remains nontrivial. To characterise biological intelligence, one must commit to a class of generic algorithms and elucidate the types of algorithms shaped by evolution.

To address these issues, the present work characterises biologically plausible algorithms arising naturally from evolution. Mathematical analyses reveal neural implementations of Turing machines—through neural activity and synaptic plasticity—and show that these networks perform variational Bayesian inference of external Turing machines' states. Additionally, the emergent mechanisms are explained through evolutionary perspectives. Bayes-optimal inference and decision making under a suitable generative model emerge via Bayesian model selection, inherent in natural selection. The paper is concluded by discussing possible neuronal mechanisms that realise the self-organising implementation of universal machines.

**RESULTS**

**Equivalence between neural networks and Bayesian inference**

First, the equivalence between canonical neural networks and variational Bayesian inference was revisited to extend it to a framework including Turing machines in the next section. Previous works have derived canonical neural networks—which exhibit certain biological plausibility—from realistic neuron models [2–4] through approximations [20]. In this work, a biological agent is defined as a canonical neural network comprising a two-layer recurrent network of rate-coding neurons (**Fig. 1a**, bottom). Upon receiving sensory inputs $o_t = (o_{t1}, \dots, o_{tN_o})^\mathrm{T}$, the middle ($x_t$) and output ($y_t$) layer neurons generate neural activities $x_t = (x_{t1}, \dots, x_{tN_x})^\mathrm{T}$ and $y_t = (y_{t1}, \dots, y_{tN_y})^\mathrm{T}$, respectively. When $x_t$ and $y_t$ have a bounded range, their values can be rescaled within the range of 0 to 1. By defining the autonomous states $u_t = \{x_t, y_t\}$, their dynamics are given by the following differential equation:

$$\dot{u}_t \propto -\mathrm{sig}^{-1}(u_t) + f(u_{t-1}, o_t) \tag{1}$$

where $\mathrm{sig}^{-1}$ indicates the inverse of the sigmoid function $\mathrm{sig}(\cdot) = 1/(1 + e^{-(\cdot)})$ and $f$ is an arbitrary vector function that maps $u_{t-1}$ to $u_t$ given $o_t$. The network's internal states involve synaptic weight matrices $\omega = \{W, K, V\}$ and firing thresholds $\phi = \{\phi^x, \phi^y\}$ that parameterise $f$, where $W$ and $K$ denote feedforward and recurrent connections to the middle layer, while $V$ denotes connections to the output layer. Actions $\delta_t = (\delta_{t1}, \dots, \delta_{tN_\delta})^\mathrm{T}$ are then sampled from $y_t$.

This model facilitates the identification of implicit Hamiltonian and Helmholtz energy in neural networks through reverse engineering [19–21]. For simplicity, the environment is expressed as a discrete state space, where $o_t$ and $\delta_t$ are vectors of binary values. A time series or path of observations is denoted as $o_{1:t} = \{o_1, \dots, o_t\}$, and the internal states of the system—including the paths of states and parameters—are denoted as $\varphi$. Without the loss of generality, equation (1) can be cast as a gradient descent on the nonsteady-state Helmholtz energy [21]:

$$\mathcal{A}[\pi(\varphi), o_{1:t}, \xi] = \langle \mathcal{H}_\xi(o_{1:t}, \varphi) + \ln \pi(\varphi) \rangle_{\pi(\varphi)} \qquad (2)$$

where $\mathcal{H}_\xi$ denotes the Hamiltonian characterised by gene $\xi$, $\langle \cdot \rangle_{\pi(\varphi)}$ is the expectation over internal state distribution $\pi(\varphi)$, and the fixed inverse temperature $\beta = 1$ is adopted for simplicity. The minimisation of $\mathcal{A}$ yields the steady-state distribution $\pi(\varphi)$. Following the treatment in previous works [19,20], the explicit form of $\mathcal{A}$ can be reverse engineered by computing the integral of the right-hand side of equation (1) with respect to $u_t$:

$$\mathcal{A}[\pi(\varphi), o_{1:t}, \xi] = \sum_{\tau=1}^{t} \begin{pmatrix} u_t \\ \overline{u_t} \end{pmatrix}^{\mathrm{T}} \left\{ \ln \begin{pmatrix} u_t \\ \overline{u_t} \end{pmatrix} - \begin{pmatrix} f_1(o_t, u_{t-1}) \\ f_0(o_t, u_{t-1}) \end{pmatrix} \right\} + \mathcal{C} \qquad (3)$$

where $f_1$ and $f_0$ are functions that satisfy $f_1 - f_0 \equiv f$, $\overline{u_t} = \vec{1} - u_t$ is the sign-flipped $u_t$ centred on 1/2, $\vec{1}$ is a vector of ones, and $\mathcal{C}$ denotes the integration constant, which is tuned to satisfy $\int e^{-\mathcal{A}} do_{1:t} = 1$ in the steady state. The distribution $\pi(\varphi)$ is encoded or parameterised by a set of internal variables $\{u_{1:t}, \omega, \phi\}$. Owing to construction, the gradient descent on $\mathcal{A}$ with respect to $u_t$, i.e., $\dot{u}_t \propto -\partial_{u_t} \mathcal{A}$, yields equation (1). Moreover, all synapses are updated by $\dot{\omega} \propto -\partial_\omega \mathcal{A}$, which results in Hebbian plasticity with a homeostatic term. In particular, $V$ exhibits three-factor Hebbian plasticity [31–33] mediated by neuromodulators $\Gamma_t = \left( \Gamma_{t1}, \ldots, \Gamma_{tN_\gamma} \right)^{\mathrm{T}}$ that encode risk [20]. Other variables are also updated by $\dot{\phi} \propto -\partial_\phi \mathcal{A}$. The explicit forms of Helmholtz energy and update rules are provided in **Fig. 1b** and the Methods section.

Crucially, according to the complete class theorem [34–36], the admissible decision rules that minimise $\mathcal{A}$ can be cast as Bayesian inferences under at least one generative model with prior beliefs. Variational Bayesian inference updates prior beliefs about external milieu states $P_m(\vartheta)$ to the corresponding approximate (or exact) posterior beliefs $Q(\vartheta)$ based on $o_{1:t}$ (**Fig. 1a**, top left), wherein variational free energy, or equivalently, the negative of evidence lower bound (ELBO)

$$\mathcal{F}[Q(\vartheta), o_{1:t}, m] = \langle -\ln P_m(o_{1:t}, \vartheta) + \ln Q(\vartheta) \rangle_{Q(\vartheta)} \qquad (4)$$

is a standard cost function [37]. Such inferences rest upon a generative model $P_m(o_{1:t}, \vartheta)$ that expresses the agent's hypothesis about how sensory inputs $o_{1:t}$ are generated from external milieu states $\vartheta$, where subscript $m$ explicates the underlying model structure. The gradient descent on $\mathcal{F}$ furnishes Bayesian belief update rules and their fixed point provides posterior belief $Q(\vartheta)$, which approximates (or may be exactly equal to) the solution of Bayes' theorem, $P_m(\vartheta|o_{1:t})$. The same strategy can be adopted to infer the optimal actions that minimise risk (or expected free energy) in the future, which is referred to as active inference [38–40].

In essence, the complete class theorem [34–36] and a series of recent works [19–21] suggest the existence of an $\mathcal{F}$ equivalent to $\mathcal{A}$, that is, $\mathcal{A} \equiv \mathcal{F}$, where $\pi(\varphi)$ encodes $Q(\vartheta)$ and the Hamiltonian corresponds to a genetically encoded generative model, $\mathcal{H}_\xi(o_{1:t}, \varphi) \equiv -\ln P_m(o_{1:t}, \vartheta)$. Specifically, previous works have constructively shown that the dynamics of canonical neural networks formally correspond to Bayesian belief updates under a class of factorial POMDPs [19,20]. This is highlighted in **Fig. 1b**, which depicts one-to-one correspondences between components of $\mathcal{A}$ and $\mathcal{F}$. Here, $\mathcal{F}$ provides an upper bound of the surprise (improbability) of sensory inputs, $\mathcal{F} \geq -\ln P_m(o_{1:t})$, where $P_m(o_{1:t}) = \int P_m(o_{1:t}, \vartheta)d\vartheta$ denotes the marginal likelihood. Thus, $\mathcal{A}$ upper bounds the surprise as well, as $\mathcal{A} \geq -\ln P_m(o_{1:t})$, where the equality approximately holds in the steady state. This notion indicates that variational Bayesian inference and surprise minimisation are inherent properties of canonical neural networks.

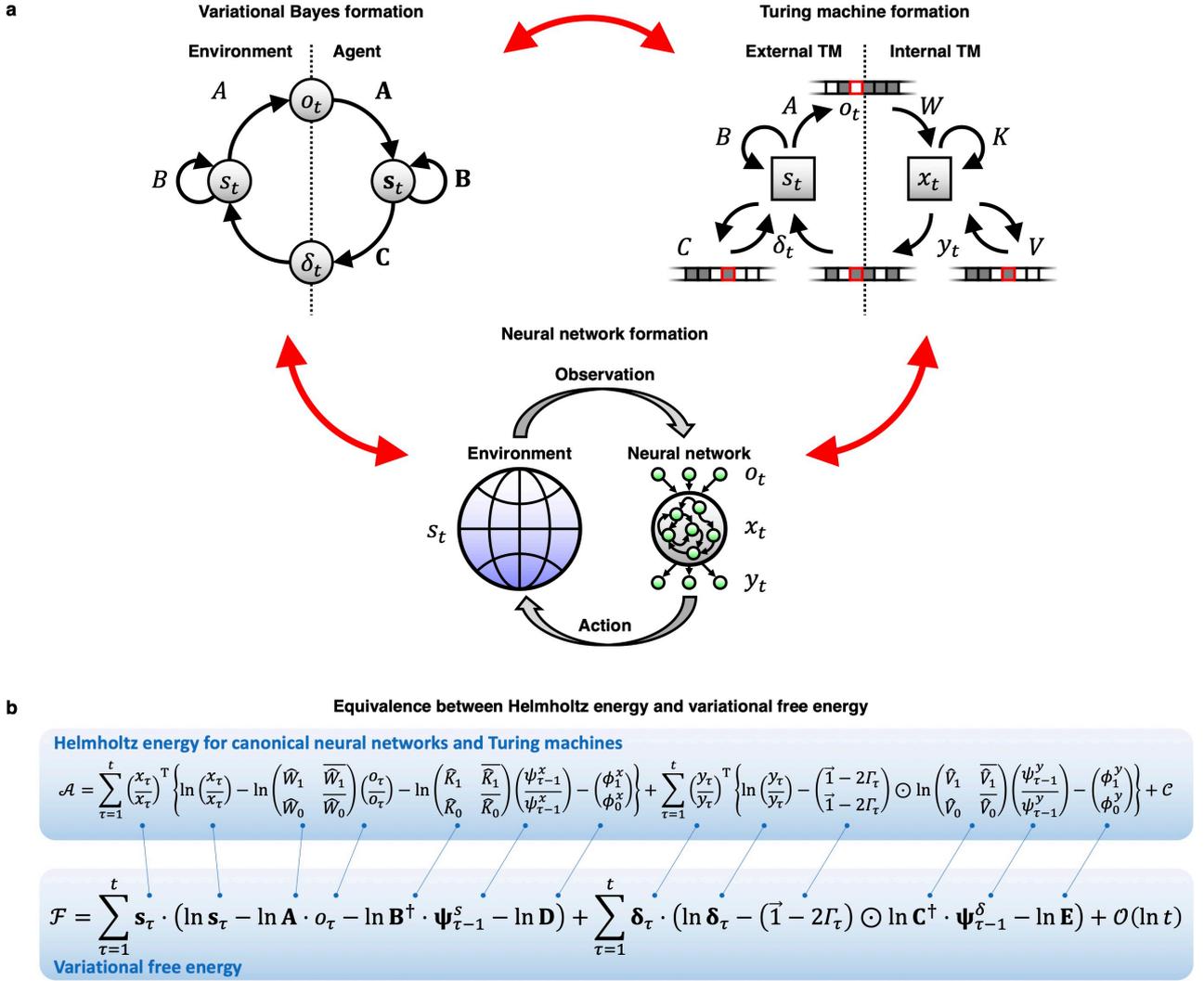

**Fig. 1. Schematic of variational trinity for biological intelligence. a.** Three theoretical categories in neuroscience—dynamical systems, Bayesian inference, and algorithms—can be integrated under the variational principle. Bottom: Dynamical system perspective, in which the activity of neurons is modelled as differential equations. The environment comprises hidden states ($s_t$) and sensory inputs ($o_t$), while a canonical neural network comprises neural activities ($x_t$) and action ($y_t$). The dynamics of neural activity and plasticity are expressed as a gradient decent on Helmholtz energy $\mathcal{A}$. Top left: Brain as a Bayesian agent. The brain is considered to perceive the environment through Bayesian inference based on past experiences. This can be formulated by minimising variational free energy $\mathcal{F}$. The posterior belief that minimises $\mathcal{F}$ provides the best guess about the external milieu states. In terms of the POMDP, a set of external variables is denoted as $\vartheta =$

$\{\delta_{1:t}, s_{1:t}, A, B, C, D, E\}$. The variables in bold (e.g., $\mathbf{s}_t$) denote the posterior beliefs about the corresponding variables in non-bold italics (e.g., $s_t$). Top right: The algorithmic perspective of the brain expressed as a Turing machine (TM). The Turing machine (internal TM) comprises a finite automaton ($x_t$) and memory or tape ($V$). Provided with automaton states $x_{t-1} \in \mathcal{X}$ and memory readout $y_{t-1} \in \mathcal{Y}$, the Turing machine update states $x_t \in \mathcal{X}$, write information in memory $\Gamma_t \in \mathcal{W}$, and move header position $\psi_t^y \in \mathcal{M}$, which is expressed as a transition mapping, $\mathcal{X} \times \mathcal{Y} \mapsto \mathcal{X} \times \mathcal{W} \times \mathcal{M}$. The environment can also be expressed as a Turing machine (external TM). **b.** Equivalence between energies for canonical neural networks, variational Bayesian inference, and differentiable Turing machines, depicted by one-to-one correspondences of their components. Here, $\overline{x_t} = \vec{1} - x_t$ denotes the sign-flipped neural activity, $\widehat{W}_1 = \text{sig}(W_1)$ denotes the sigmoid function of synaptic weights, $h_1^x = \ln \widehat{W}_1 + \phi_1$ denotes the firing threshold, $\odot$ denotes the Hadamard product operator, $\boldsymbol{\psi}_t^s$ and $\boldsymbol{\psi}_t^\delta$ are basis functions, and $\mathbf{B}^+ = \mathbf{B}^\text{T}\text{diag}[\mathbf{D}]^{-1}$ and $\mathbf{C}^+ = \mathbf{C}^\text{T}\text{diag}[\mathbf{E}]^{-1}$ are inverse mappings (see Methods for details). A differentiable Turing machine can be derived as gradient descent on the Helmholtz energy that is shared with canonical neural networks (see main text). Therefore, the three perspectives can be unified in terms of the variational principle—a notion referred to as variational trinity.

**Equivalence between neural networks, Bayesian inference, and Turing machines**

Here, a class of Turing machines are shown to be equivalent to canonical neural networks and variational Bayesian inference under a class of POMDPs. A Turing machine, denoted as $\text{TM}(\cdot)$, is a foundational model in computational science. It comprises a finite-state machine or automaton ($x_t$) for computations and a sufficiently long tape or large memory ($V$) to retain information [1], enabling the expression of arbitrary algorithms (**Fig. 1a**, top right). For each timestep $t > 0$, the Turing machine reads memory value $y_{t-1}$ and observation $o_t$, updates the automaton states $x_t$, writes a new value to the memory depending on $\Gamma_t$, and moves header position $\psi_t^y$, defined formally as

$$\{x_t, \Gamma_t, \psi_t^y\} = \text{TM}(x_{t-1}, y_{t-1}, o_t) \tag{5}$$

Because Turing machines are a family of discrete state space models, equation (5) can be expressed in the form of POMDPs using categorical distributions, where $\Gamma_t$ and $\psi_t^y$ are viewed as deterministic functions (see Methods for details). Moreover, the dynamics of $u_t = \{x_t, y_t\}$ that converge to equation (5) can be formulated as equation (1), which can be read as a differentiable Turing machine [26]. Such Turing machines and the corresponding Helmholtz energy $\mathcal{A}$ are provided by substituting $\text{sig}(f(u_{t-1}, o_t)) \equiv \text{TM}(u_{t-1}, o_t)$ into equations (1) and (3), respectively, where the fixed point ($\partial_{u_t}\mathcal{A} = 0$) yields equation (5). Hence, arbitrary algorithms can be derived as the gradient descent on the implicit $\mathcal{A}$.

Remarkably, canonical neural networks and differentiable Turing machines share an identical Helmholtz energy as shown in **Fig. 1b** and in the Methods section, which allows differentiable Turing machines to be implemented within canonical neural networks. In this expression, the middle ($x_t$) and output ($y_t$) layer neurons are cast as the finite-state machine and memory readout, respectively, where $y_t$ can be viewed as a mental action that changes the network's internal states [41]. The synaptic weights in the middle ($K$) and output ($V$) layers encode the transition mapping and memory, respectively. Memory writing occurs through quick synaptic plasticity in $V$ modulated by neuromodulator $\Gamma_t$, which is derived from the Helmholtz energy minimisation. This plasticity rule retains the current memory value (or strategy) $y_{ti}$ for subsequent use when the risk is low ($\Gamma_{ti} = 0$). Conversely, it forgets the current value and writes the opposite value $\overline{y_{ti}}$ at the current memory location when the risk is high ($\Gamma_{ti} = 1$). These rules can be expressed as the product of activities of pre ($\psi_{t-1}$) and post ($y_t$) synaptic neurons and neuromodulators ($\vec{1} - 2\Gamma_t$) in the form of a three-factor Hebbian rule [31–33]. With this architecture, the network can store and execute programs such as 'for' loops or 'if' or 'go to' statements in a certain memory location. This minimal configuration enables the computation of arbitrary algorithms.

Furthermore, these activities and plasticity can be read as variational Bayesian inferences under a class of factorial POMDPs, owing to the established equivalence, which enables the deployment of strong variational Bayesian inference tools to explain the characteristics of canonical neural networks implementing Turing machines. The one-to-one correspondence of their components is summarised in **Table 1**. The formal definition of a Turing machine in terms of a probabilistic generative model is explained in the Methods section. In essence, the dynamics of canonical neural networks can be read as performing a Bayesian inference of external Turing machines, wherein $x_t$, $y_t$, and $\omega$ encode the posterior belief about hidden states $s_t$, memory readout $\delta_t$, and parameters $\theta$, respectively. Particularly, when the environment is expressed by deterministic mappings, the states and parameters of canonical neural networks tend to take either 0 or 1 to match those of external milieu as learning progresses. Owing to this property, biological agents can recapitulate algorithms existing in the external milieu and imitate their computations within neural networks in a self-organising manner.

These propositions were corroborated using numerical simulations, demonstrating that canonical neural networks can infer the computational processes of foundational algorithms, such as adders, based on noisy observations (**Fig. 2a**). Here, the external Turing machine comprises hidden states ($s_t$) and memory ($C$), which involve the information of 16-dimensional binary numbers and their sum, respectively. The sensory inputs ($o_t$) comprised sets of MNIST handwritten digits (0 or 1) representing 16-digit binary numbers that were generated based on $s_t$. Learning continued for four sessions (2,048 steps). On receiving these inputs, the canonical neural networks were able to infer hidden states (**Fig. 2b**) and add the binary numbers iteratively to synaptic weights $V$ (**Fig. 2c**). The matching between the numbers encoded in $C$ and $V$ indicates the success of the Bayesian inference of the hidden states and memory of the external Turing machine (**Fig. 2d**). Over sessions, the error in imitating the memory dynamics of the external Turing machine was decreased (**Fig. 2e**). Moreover, the networks could learn the parameters of the external Turing machines (**Fig. 2f**), thereby recapitulating their programs within the network's synaptic weights.

In summary, the equivalence between canonical neural networks, variational Bayesian inference, and differentiable Turing machines was established. The dynamics of canonical neural networks can be read as Bayesian belief updates that entail the agent's internal states encoding posterior beliefs about the external Turing machine. These results imply the intrinsic universality of canonical neural networks with mental actions and fast modulated plasticity, enabling the inference of external algorithms in a biologically plausible manner. However, these inferences rely on genetically encoded generative models, in which arbitrarily selected genes are suboptimal for a given environment. Thus, the remainder of this paper addresses the types of algorithms or generative models that emerge evolutionarily through interactions with the environment.

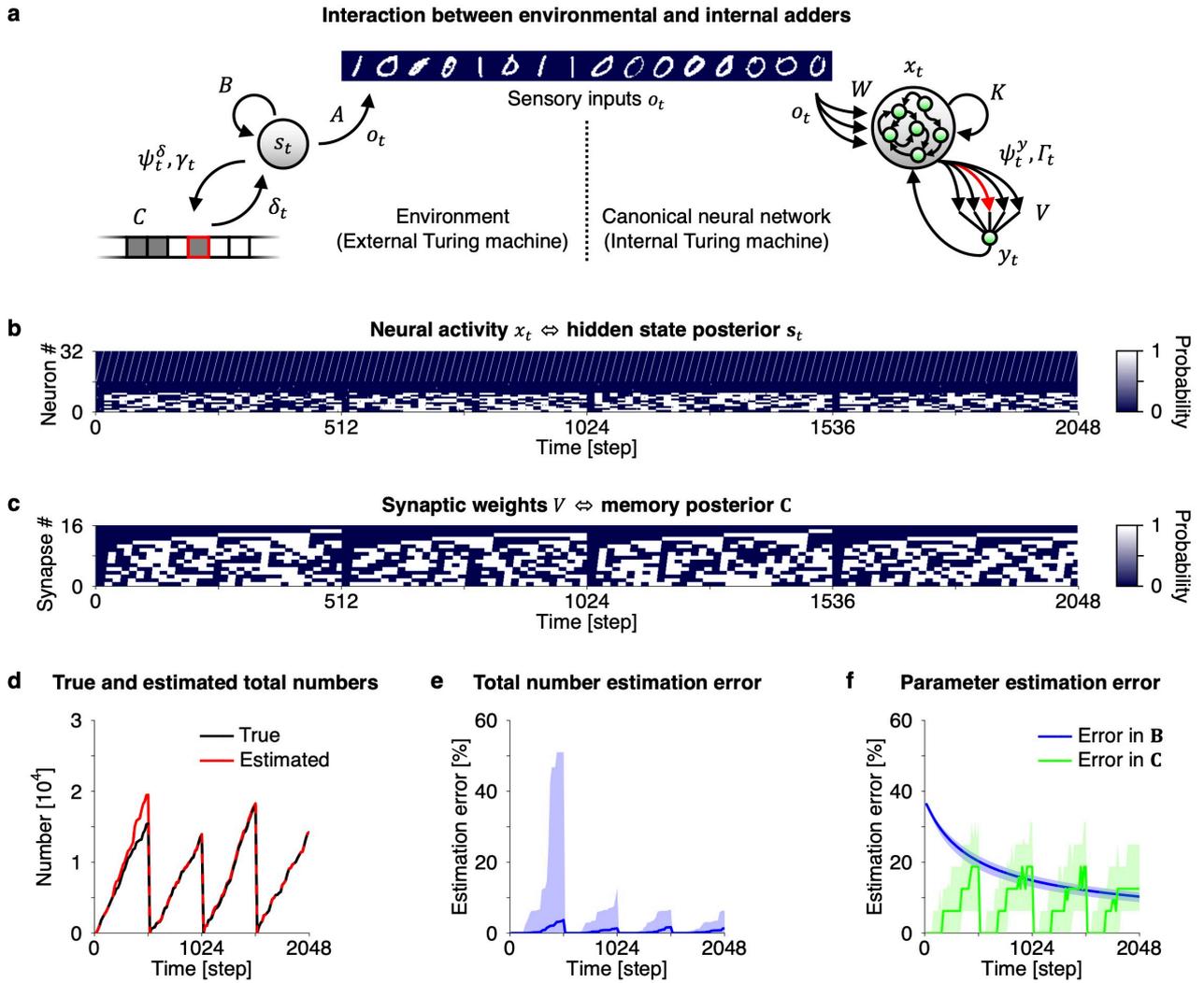

**Fig. 2. Canonical neural network implementation of adders. a.** Schematic of interaction between external adder and canonical neural network. The environment comprises sensory inputs ($o_t$), hidden states ($s_t$), readout ($\delta_t$), header position ($\psi_t^\delta$), writing ($\gamma_t$), memory ($C$), and automaton parameters ($\theta$) that encode transition mappings. Sensory inputs comprise a sequence of handwritten 16-digit binary numbers and a 16-step-cycle counter to count the passage of time. An example sensory input digit pattern is shown on the top. Each handwritten digit comprises a 28 × 28 image taking values of 0 or 1, and the counter input comprises 256 states to represent 16 steps. Thus, sensory inputs are 12,800-dimensional binary vectors. The binary numbers encoded in the hidden states are added in the memory. **b.** Transition of neural activity ($x_t$) that encodes the hidden state posterior ($\mathbf{s}_t$). Hidden states comprise a 32-dimensional binary vector, in which half

the states (16) correspond to 16-digit binary numbers and the other half represent the counter that iterates 16 step sequences. The neural activity ($x_t$) matches the true hidden states ($s_t$), indicating the success of canonical neural networks performing Bayesian inference of external Turing machine's states. **c.** Transition of synaptic weights ($V$) that encode the posterior expectation of memory states (**C**). The values of $V$ express the summation of numbers encoded by hidden states. **d.** Matching between true and estimated binary numbers. The posterior expectation of the memory is in good agreement with the external memory values. **e.** Total number estimation error decreasing with time. **f.** Error in estimating parameters of external adders. The estimation error of the $B$ matrix decreases continuously with time, and that of the memory ($C$) decreases with sessions. Hence, canonical neural networks can implement an adder and infer the external adder's states in a self-organising manner. In (e) and (f), lines and shaded areas represent median values and areas between the first and third quartiles obtained from 100 simulations.

**Helmholtz energy minimisation derives natural selection and active Bayesian model selection**

In this section, evolution is characterised in terms of Helmholtz energy minimisation and Bayesian model selection (**Fig. 3a**). Evolution is driven by the differential survival of individuals according to their fitness and survivability, which determine the transmissibility of their distinct genetic blueprints. Each biological agent is encoded by an $N_\xi$-dimensional gene $\xi \in \{0,1\}^{N_\xi}$, which characterises the network architecture and implicit algorithm, including network size. Natural selection is characterised by the reproduction rate $\rho(o_{1:T}, \xi)$ that determines the expected number of offspring per an individual as a function of sensory input sequence $o_{1:T}$ and gene $\xi$, where $T$ denotes the lifetime duration. The gene distribution of the current generation is denoted as $n(\xi)$, and the offspring distribution is given as

$$N(o_{1:T}, \xi) = \rho(o_{1:T}, \xi) P(o_{1:T}|\xi) n(\xi) \quad (6)$$

This indicates that the number of offspring that equips $\xi$ and observes $o_{1:T}$ is determined based on the product of $\rho(o_{1:T}, \xi)$, $n(\xi)$, and conditional probability $P(o_{1:T}|\xi)$. In the presence of genetic mutation during natural selection, $n(\xi)$ in equation (6) may be replaced with $n'(\xi) = v(\xi) * n(\xi)$, where $v(\xi) *$ indicates the convolution with mutation probability $v(\xi)$. Equation (6) selects a favourable environment and gene distribution for the next generation, which is the driving force of evolution. The gene is selected more frequently if the corresponding phenotype receives preferable sensory inputs $o_{1:T}$ associated with a higher reproduction rate. After sufficient evolution, only genes that maximise the number of offspring survive, resulting in the marginal gene distribution $N(\xi) = \sum_{o_{1:T}} N(o_{1:T}, \xi)$ with sharp peaks at the optimal genes. However, this classical view does not specify the characteristics of the emergent phenotypes. Thus, it is necessary to phenotype the synthetic species (i.e., algorithms) encoded by the selected genes in relation to the given environment.

Remarkably, natural selection can be derived from variational principles at the species or population level. While each individual minimises the Helmholtz energy $\mathcal{A}$ under a particular $o_{1:T}$, the entire species experiences a sufficient number of various $o_{1:T}$ generated from the generative process $P(o_{1:T}, \vartheta)$, which facilitates the computing of the expectation over $o_{1:T}$. By extending equation (3), the ensemble Helmholtz energy for the entire species is defined as follows:

$$\overline{\mathcal{A}}[\pi(o_{1:T}, \xi)] = \langle \mathcal{A}[\pi(\varphi), o_{1:T}, \xi] - \ln \rho(o_{1:T}, \xi) - \ln n(\xi) + \ln \pi(o_{1:T}, \xi) \rangle_{\pi(o_{1:T}, \xi)} \quad (7)$$

where $\pi(o_{1:T}, \xi)$ denotes the joint distribution of $o_{1:T}$ and $\xi$ for the offspring generation and $\langle \cdot \rangle_{\pi(o_{1:T}, \xi)}$ is the expectation over $\pi(o_{1:T}, \xi)$. During the lifetime, each synthetic species optimises $\pi(\varphi)$ by interacting with the environment, followed by determination of the offspring distribution $\pi(o_{1:t}, \xi)$, each of which can be read as an individual adaptation and natural selection, respectively. As noted above, $\mathcal{A}[\pi(\varphi), o_{1:T}, \xi]$ converges to $-\ln P_m(o_{1:t})$ in the steady states. From the variational method, solving the fixed point $\delta\overline{\mathcal{A}} = 0$ with respect to $\pi(o_{1:t}, \xi)$—after the convergence of $\pi(\varphi)$ to the steady state—yields the offspring distribution:

$$\pi(o_{1:t}, \xi) = \frac{1}{Z}\rho(o_{1:t}, \xi)P_m(o_{1:t})n(\xi) \tag{8}$$

where $Z = \sum_{o_{1:t},\xi} \rho(o_{1:t}, \xi)P_m(o_{1:t})n(\xi)$ denotes the partition function. When genetically encoded generative models have sufficient representation capacity, the marginal likelihood $P_m(o_{1:t})$ can be viewed as a good approximation of the conditional likelihood $P(o_{1:t}|m)$. Moreover, because $\xi$ encodes model structure $m = m(\xi)$, $P(o_{1:t}|m) = P(o_{1:t}|\xi)$ holds by construction. Thus, equation (8) is asymptotically equivalent to the offspring distribution derived from natural selection (equation (6)) up to the normalisation. Moreover, $Z$ formally corresponds to the total number of offspring, which is consistent with previous work [42–44]. Further details are provided in the Methods section. Hence, although each individual does not know $\rho(o_{1:t}, \xi)$ or $n(\xi)$ directly, the variational principle enables the optimisation of $\pi(o_{1:t}, \xi)$ through the interaction between the species and the environment.

According to the equivalence [19–21], a variational free energy $\overline{\mathcal{F}}$ corresponding to $\overline{\mathcal{A}}$ exists, which yields the inequality of evolution:

$$\overline{\mathcal{A}}[\pi(o_{1:T}, \xi)] \equiv \overline{\mathcal{F}}[Q(o_{1:T}, m)] \geq -\ln Z \tag{9}$$

This indicates that $\overline{\mathcal{A}}$ upper bounds the negative logarithm of the total offspring number $Z$. The equality holds at the steady state, which is obtained by substituting equation (8) into equation (7). Emergent synthetic species can yield adaptive sentient behaviours that maximise the reproduction rate. When the gene mutation probability is sufficiently small, agents with other suboptimal genes—which have biased generative modes and lower survival and reproduction probabilities—are asymptotically eliminated after sufficient evolution. This results in the emergence of sharp peaks in gene distribution. Variations in $\xi$ may exist if these are unrelated to optimisation. Another gene distribution may appear when the gene mutation probability is large [45].

It is crucial that equation (9) also holds true for the variational free energy $\overline{\mathcal{F}}$, where $Q(o_{1:T}, m)$ denotes the posterior belief about the offspring's observations and genes. Minimising $\overline{\mathcal{F}}$ ($\equiv \overline{\mathcal{A}}$) furnishes an active Bayesian model selection that selects desirable sensory inputs $o_{1:t}$ for the next generation. An optimal gene that minimises $\overline{\mathcal{A}} \equiv \overline{\mathcal{F}}$ simultaneously minimises $-\ln Z$, by which the agent attains the generative model $P_m(o_{1:T}, \vartheta)$ that best recapitulates the true generative process of external milieu $P(o_{1:T}, \vartheta)$, known as Hamiltonian matching [21]. This occurs because the offspring number is maximised most efficiently—and thus the gene distribution reaches a steady state—when $P_m(o_{1:T}, \vartheta)$ best matches $P(o_{1:T}, \vartheta)$. Hence, natural selection derived from the Helmholtz energy minimisation acts as active Bayesian model selection that optimises the generative model of agents. These observations are consistent with the outcomes in previous works that applied the free-energy principle to the explanation of evolutionary systems [46–49].

In particular, when the external milieu comprises the considered class of Turing machines, an optimal generative model is a canonical neural network. Thus, ensemble Helmholtz energy minimisation over generations makes synthetic species have evolved to equip canonical neural networks. These networks then make Bayesian inference of the states of external Turing machines in a self-organising manner.

These propositions were corroborated by numerical simulations of evolution. The external milieu was characterised by adders considered in **Fig. 2**. Biological agents were characterised by the genes ($\xi$) and formulated by equation (1). The simulation comprised a cycle of adaptation and evolution (**Fig. 3a**). The genes were updated through genetic mutation at the natural selection stage, in which agents with a higher accuracy of estimating the summation of input numbers exhibited a greater reproduction rate. At the beginning of evolution, neural networks employed the Helmholtz energy characterised by randomly selected genes, which derived suboptimal algorithms or strategies. The evolutionary process selects a generative model and risk that maximise the reproduction. As expected, only species with the optimal gene survived after sufficient generations, and the final gene distribution exhibited a sharp peak at the optimal gene that encodes an apt generative model (**Fig. 3b**). These synthetic species evolved to exhibit Bayesian belief updating under the optimal (i.e., unbiased) prior belief and yielded superior performance compared to other species with suboptimal genes. This was confirmed by monotonic decrease of memory state estimation errors over generations (**Fig. 3c**). Consequently, the surviving species could implement an adder in their network architecture and estimate the summation of input binary numbers (**Fig. 3d**).

In summary, ensemble Helmholtz energy minimisation can be used to derive natural selection that maximises the total number of offspring. The natural selection functions as active Bayesian model selection, which facilitates emergent species to best recapitulate the given environment and exhibit Bayes-optimal decision making. This notion renders evolution explainable in terms of Bayesian inference and generative models, facilitating the universal characterisation of biological intelligence emerging from evolution.

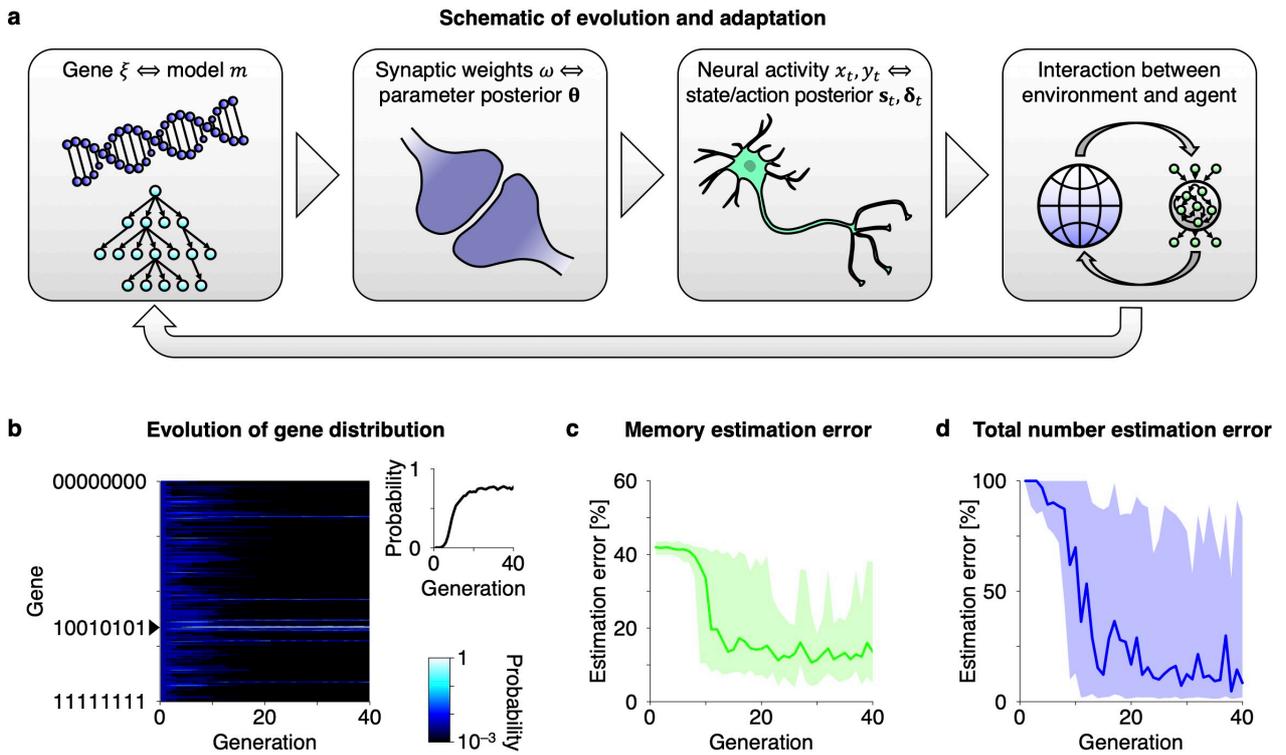

**Fig. 3. Evolutionary emergence of neural networks implementing Turing machines. a.** Flow diagram of the evolutionary scheme. Natural selection optimises the genes ($\xi$) that determine the form of the Helmholtz energy, whereas synaptic plasticity (i.e., learning) and neural activity (i.e., inference) update the posterior beliefs about parameters ($\theta$) and autonomous states ($s_t, \delta_t$), respectively. Agents consequently engage in actions that minimise the risk associated with future outcomes. These processes optimise across various time scales and hierarchies, all fundamentally based on Helmholtz energy minimisation. **b.** Emergence of neural networks encoding adders. Environment is characterised by adders same as in **Fig. 2**. The reproduction rate is determined by the estimation accuracy of summation of the input numbers. The heat map shows evolutionary change in gene distribution. Each generation consists of 100 agents. The initial condition corresponds to the suboptimal, randomly selected generative model. A cycle of adaptation and evolution facilitates active Bayesian model selection, by which the gene that encodes an optimal generative model is selected, namely, $\xi = (1,0,0,1,0,1,0,1)^{\mathrm{T}}$ in this setting. Top right panel shows the trajectory of the probability of the optimal gene. **c.** Errors in estimating memory states. The synaptic strengths (*V*) encode the posterior belief about memory (*C*). **d.** Decrease of total number estimation error with the increase in generation. In (c) and (d), lines and shaded areas represent median values and areas between the first and third quartiles obtained with 100 agents.

**Canonical neural networks can implement universal Turing machines**

Finally, the possibility of canonical neural networks becoming universal machines is addressed. Some Turing machines can emulate an algorithm implemented by other Turing machines within a single architecture by storing programs in memory. These are referred to as universal Turing machines [1], and can be constructed using a small circuit, such as that with $N_x = 2$ states and 18 memory values [50] or $N_x = 18$ states and 2 memory values [51]. This implies that even relatively small neural networks can function as universal machines.

These universal machines can be straightforwardly implemented using a POMPD with two headers (actions) and memories. In the POMDP formation, the state transition is determined based on a set of previous states ($s_{t-1}$), memory readout ($\delta_{t-1}$), and transition matrix ($B$), as $s_t = B(s_{t-1} \otimes \delta_{t-1})$ using a Kronecker product. This constructs an algorithm. Conversely, the product of new $\delta'_{t-1}$ and $B'(\cdot)$ can be used to construct transition mapping $B'(\delta'_{t-1})$ that follows the program encoded in $\delta'_{t-1}$. Then, with the second memory readout $\delta_{t-1}$, transition mapping $s_t = B'(\delta'_{t-1})(s_{t-1} \otimes \delta_{t-1})$ is constructed, with which agents can emulate the aforementioned algorithm when $\delta'_{t-1}$ satisfies $B'(\delta'_{t-1}) = B$. This is a construction of a universal Turing machine, which enable the computation of arbitrary transitions of $s_t$ according to $\delta'_t$ and $\delta_t$.

Owing to the equivalence, canonical neural networks with two mental actions can implement universal Turing machines in a straightforward manner. Diverse programs can be executed with a single neural network by reading them from the output-layer synaptic weights. Using this architecture, canonical neural networks can learn to infer the states of multiple Turing machines in the environment. Selecting an appropriate transition mapping by moving a header position is associated with attentional switch [41,52], which may be encoded by neuromodulators such as dopaminergic neurons [53].

The simulation environment included 10 Turing machines, with each comprising 10 states ($s_t \in \{0,1\}^{10}$) and a memory of length 10 ($C \in \{0,1\}^{1 \times 10}$) as well as randomly generated transition rules ($B \in \{0,1\}^{10 \times 10 \times 2}$) that differ from each other (**Fig. 4a**). Agents could observe sensory inputs $o_t \in \{0,1\}^{10}$ generated by $s_t$, whereas the other variables were unobservable. For simplicity, in this simulation, the agents and external Turing machines were considered to share common memory writing rules ($\Gamma_t$) and likelihood mapping ($A$). In this environment, the Turing machines were switched at random intervals. Thus, agents were required to infer which Turing machine was generating the current input and what transition rules it employed.

The simulation results demonstrated that the canonical neural network successfully distinguished among 10 different Turing machines by accumulating evidence (**Fig. 4b**) and predicted the dynamics of external Turing machines by learning and storing their transition matrices in the memory. The header of the memory location corresponds to the model with the highest likelihood, analogous to the mixture generative model [41] and neural Turing machine [26,27] proposed in previous works. During a $10^5$-time-step session, canonical neural network successfully learned to encode 10 transition matrices within synaptic weights *V* (**Fig. 4c**). This was reliably observed in 20 simulations with different environmental settings (**Fig. 4d**). Importantly, because these transition matrices are stored individually in *V*, the agent allows to reuse the transition matrices learned in the past to make rapid predictions when receiving sensory inputs from previously experienced Turing machines. Owing to this, canonical neural networks could move the header to the memory location encoding a Turing machine that was generating the inputs, with small estimation error (**Fig. 4e**). Accordingly, errors in predicting the subsequent hidden states decreased with time (**Fig. 4f**).

In summary, canonical neural networks can biologically plausibly implement universal Turing machines using two mental actions and fast modulated synaptic plasticity. These characteristics are crucial for developing a flexible and adaptable biological intelligence. It is the virtue of canonical neural networks that they can emulate arbitrary algorithms owing to their Turing completeness or computational universality [1].

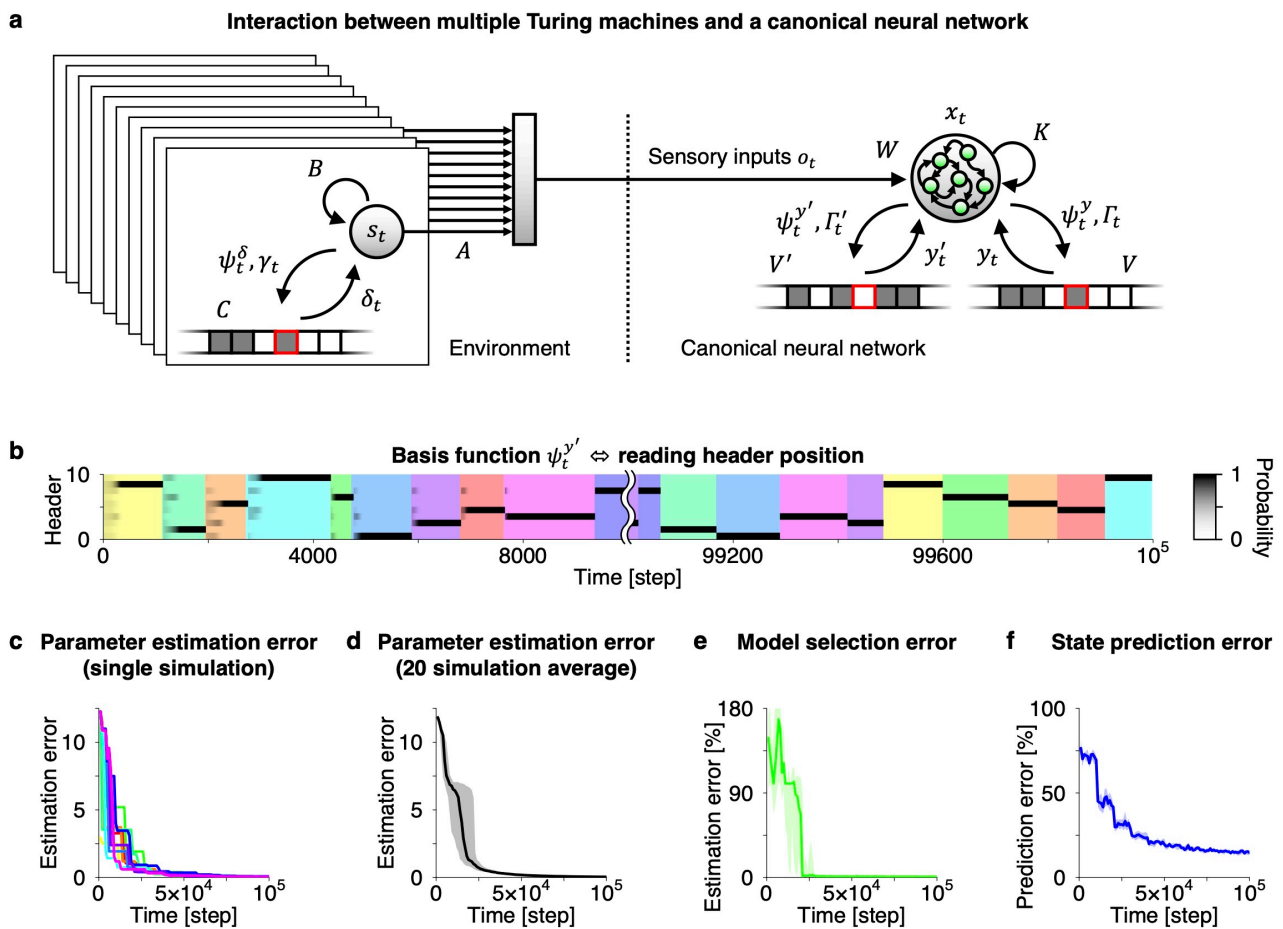

**Fig. 4. Neuronal implementations of universal machines. a.** Schematic of the interaction between 10 external Turing machines and a canonical neural network involving two mental actions. **b.** Transition of the basis function ($\psi_t^{y'}$) that encodes the reading header position. After training, the network accurately infers which the Turing machine is generating the sensory inputs. **c.** Errors in estimating the transition matrices ($B$) of the external Turing machines. Ten colour lines indicate estimation errors for 10 Turing machines in a simulation. **d.** Errors in estimating the transition matrices over 20 simulations. **e.** Errors in inferring which Turing machine is generating the input. The results indicate that the reading header ($\psi_t^{y'}$) moves to the correct position after learning. **f.** Errors in predicting subsequent hidden or automaton states ($s_{t+1}$) of the external Turing machines. In (d), (e), and (f), lines and shaded areas represent median values and areas between the first and third quartiles obtained from 20 simulations.

**DISCUSSION**

Most computational models in neuroscience have employed neural networks specialised for a specific task. Although previous works investigated artificial neural networks with external memory [26,27] and neuronal implementations of Turing machines [28–30], the emergent mechanics of Turing machines is a more delicate problem. In this work, canonical neural networks that could autonomously acquire diverse algorithms by performing Bayesian inference of external Turing machines were constructed. The roles of mental action and modulated plasticity are crucial for implementing memory reading and writing. Brain regions with sparse information representation and high plasticity, such as the hippocampus [54], may play a role of memory ($V$). Their update rules naturally emerge as a corollary of Helmholtz energy minimisation and implicit Bayesian belief updates. This is distinct from conventional neural networks that only employ a fixed generative model or finite-state machine without tape, which can only perform a specific hard-coded program. Canonical neural networks are potentially important for addressing these limitations and modelling generic intelligence.

In the considered setting, time progresses continuously as the network processes the signals. Therefore, if the process exceeds the lifetime duration, agents with suboptimal algorithms cannot produce efficient outcomes, and would be eliminated during natural selection. Thus, the surviving species may efficiently avoid the so-called hold problem.

Genes that encode programs or generative models are shaped through natural selection. Characterising the associated algorithms is essential for understanding biological intelligence. Owing to the correspondence between natural selection and active Bayesian model selection, biological agents evolve to perform active inference of the external milieu states under an apt generative model. This is attributed to the complete class theorem [34–36], which guarantees that algorithms that minimise cost functions—in the sense of admissible rules—can be read as performing Bayesian inferences. This notion lends the explainability of Bayesian inference to the characterisation of generic biological agents (i.e., algorithms) and evolution.

The synthetic species follow an asymptotic trajectory to identify generative models that best recapitulate environmental generative processes in their internal states. This notion describes the emergent mechanisms of surprise minimisation, generative models, and preference priors in neural networks—which are provided as a priori assumptions in the current formulation of the free-energy principle [17,18,38–40]—from an evolutionary perspective. The generative model implemented in the network architecture becomes accurate through natural selection, because it most effectively minimises the Helmholtz energy and thus maximises their chances of reproduction. These optimisations may be accelerated by combining local gradient descent with genetic algorithms [55].

Generally, the external milieu comprises an environment and multiple agents that adapt and evolve, prompting us to consider the coevolution of these agents. Generic dynamical systems, including multiple agents, can be described by interactions between Turing machines. An extension of the proposed framework to the coevolution of multiple agents is described in the Methods section, although details are not provided in this paper. The analyses suggest that the coevolution of multiple agents entails the emergence of generalised synchrony in agents' internal states and strategies [56,57], wherein each agent infers the algorithms of other agents. The Bayesian perspective suggests that some psychiatric disorders are caused by a break in balance in the integration of sensory information with prior beliefs [58]. Mutual inferences of Turing machines could be related to the model of psychiatric disorders associated with attenuated social intelligence. In the presence of social interactions and coevolution, the complexity of the generation model tends to increase because each agent evolves to infer and imitate algorithms of other agents. These processes lead to chaotic progressions in which the final states can vary significantly. Because of these difficulties, formal characterisations of social interaction and coevolution are beyond the scope of this paper, which is an important research direction for future work.

In summary, the equivalence between canonical neural networks, Bayesian inference, and Turing machines was demonstrated. Canonical neural networks with mental actions and modulated plasticity can implement a Turing machine in a biologically plausible manner and perform variational Bayesian inferences of external Turing machines. The Helmholtz energy minimisation at the species level involves natural selection. An appropriate network architecture (i.e., generative model) naturally emerge through active Bayesian model selection, which is implicit in natural selection. These notions offer a universal characterisation of biological intelligence emerging naturally from evolution, providing a possible explanation for the emergence of generic biological intelligence and an insight into the development of artificial general intelligence.

**METHODS**

**Turing machines in the form of POMDP generative models**

In this work, the environment is defined as a Turing machine expressed in the form of a factorial POMDP. The Turing machine comprises internal states of automaton $s_t \in \mathcal{S}$, memory $C$, readout from the memory $\delta_t \in \mathcal{D}$, header position $\psi_t^\delta \in \mathcal{M}$, memory writing $\gamma_t \in \mathcal{W}$, and transition mapping $\mathcal{S} \times \mathcal{D} \mapsto \mathcal{S} \times \mathcal{W} \times \mathcal{M}$. The initial and final (accepting) states are denoted as $s_1$ and $s_{acc}$, respectively. Memory $C$ has a finite but sufficient length, and its elements usually take a binary value. In terms of the generative model, equation (5) is characterised by categorical distributions, from which binary vector variables $s_t$, $\delta_t$, and $\gamma_t$ are sampled. The POMDP becomes a deterministic Turing machine with deterministic transition mappings, whereas it becomes a stochastic Turing machine when state transitions involve some stochasticity.

The automaton updates the hidden states $s_t$ following $P(s_t|s_{t-1}, \delta_{t-1}, B) = \text{Cat}(B\psi_{t-1}^S)$, where $\psi_t^S = \psi(x_t, y_t)$ denotes basis functions which is an arbitrary function of $x_t$ and $y_t$; for example, it is defined as a Kronecker product of $s_t$ and $\delta_t$, $\psi_t^S = s_t \otimes \delta_t$. Then, it generates (mental) actions $\delta_t$ depending on the policy or memory matrix $C$, in which $\delta_t$ formally corresponds to the memory readout that reads the memory information at position $\psi_{t-1}^\delta$. This is given as $P(\delta_t|s_{t-1}, \delta_{t-1}, C) = \text{Cat}(C\psi_{t-1}^\delta)$ using basis function or header position. The reading header position $\psi_t^\delta = \psi^\delta(s_t, \delta_t, \psi_{t-1}^\delta)$ moves depending on $s_t$, $\delta_{t-1}$, and $\psi_{t-1}^\delta$. State basis $\psi_t^S$ and reading header position $\psi_t^\delta$ are treated as dependent variables. The Turing machine further writes the new information to the memory at the current header position depending on risk $\gamma_t$, which follows $P(\gamma_{ti} = 1) = \Gamma_{ti}$, where $\Gamma_t = \Gamma(s_t, \delta_t, o_t)$. In the writing phase, $C\psi_{t-1}^\delta$ is retained when $\gamma_t = 0$ or inverted when $\gamma_t = 1$. The output of the Turing machine becomes the sensory input $o_t$ for biological agents: $P(o_t|s_t, A) = \text{Cat}(As_t)$. The sensory input can be viewed as a readout from a tape shared by the environment and agent.

Therefore, the generative model is defined as follows:

$$P_m(o_{1:t}, s_{1:t}, \delta_{1:t}, \theta, \lambda) = P(\theta)P(\lambda)\prod_{\tau=1}^{t} P(o_\tau|s_\tau, A)P(s_\tau|s_{\tau-1}, \delta_{\tau-1}, B)P(\delta_\tau|s_{\tau-1}, \delta_{\tau-1}, \gamma_\tau, C) \quad (10)$$

where $\theta = \{A, B, C\}$ denotes a set of parameters and $\lambda = \{D, E\}$ denotes a set of hyper parameters. Components of $P(\theta)P(\lambda) = P(A)P(B)P(C)P(D)P(E)$ follow Dirichlet distributions. These parameters have prior beliefs that persist with the values learned in the past, whereas $C$ changes quickly compared to other parameters owing to an effectively small Dirichlet count. To minimise the risk, the $C$'s posterior belief is updated using a counterfactual generative model defined previously [20], in which $P(\delta_\tau|s_{\tau-1}, \delta_{\tau-1}, \gamma_\tau, C) = \text{Cat}(C\psi_{\tau-1}^\delta)$ for $\gamma_\tau = 0$ while $P(\delta_\tau|s_{\tau-1}, \delta_{\tau-1}, \gamma_\tau, C) \propto \text{Cat}(C^{\odot-1}\psi_{\tau-1}^\delta)$ for $\gamma_\tau = 1$ using the Hadamard power of $C$. The initial values of $C$ may follow a flat prior of 0.5, which expresses a white space character. Notably, POMPDs with a factorial structure are considered herein but a simplified notation is adopted; for more details, please refer to previous work [19,20]. A set of external milieu variables is denoted as $\vartheta = \{s_{1:t}, \delta_{1:t}, \theta, \lambda\}$. The generative model $P_m(o_{1:t}, \vartheta)$ may or may not be equal to the true generative process of the external milieu, $P(o_{1:t}, \vartheta)$, where model structure $m$ is optimised during the evolution.

In essence, the considered class of POMDPs is cast as Turing machines when the $C$ matrix can be read as a sufficiently long tape (or memory), in which the column index of $C$ corresponds to the memory address. A sequential access header $\psi_t^\delta$ only moves to the right or left cell from the current position, which is sufficient for Turing machines. However, random-access header $\psi_t^\delta$ that freely moves to any address location is more useful for practical configurations, as considered in previous works [26,27]. Generally, multiple actions $\delta_t$ may separately express memory readout (i.e., mental action) and direct feedback responses to the environment.

**Variational free energy minimisation**

Variational free energy (equation (4)) provides an upper bound of the surprise $-\ln P_m(o_{1:t})$. Solving the fixed point of the $\mathcal{F}$'s variation $\delta \mathcal{F} = 0$ provides the approximate posterior belief $Q(\vartheta)$, which minimises $\mathcal{F}$ and results in $\mathcal{F} \simeq -\ln P_m(o_{1:t})$. When the environment is expressed as POMDPs, posterior expectation $\boldsymbol{\vartheta} \coloneqq \mathbb{E}_{Q(\vartheta)}[\vartheta]$ or its counterpart is sufficient to approximate $Q(\vartheta)$. Thus, $\mathcal{F}$ is reduced to a function of $\boldsymbol{\vartheta}$, $F(o, \boldsymbol{\vartheta})$. Throughout the paper, bold case variables (e.g., $\boldsymbol{\vartheta}$) are used to indicate the posterior expectation of the corresponding random variable (e.g., $\vartheta$).

In this setting, an approximate posterior belief is given as follows:

$$Q(s_{1:t}, \delta_{1:t}, \theta) = Q(A)Q(B)Q(C) \prod_{\tau=1}^{t} Q(s_\tau)Q(\delta_\tau) \qquad (11)$$

where $Q(s_\tau)$ and $Q(\delta_\tau)$ are categorical distributions and $Q(A)$, $Q(B)$, and $Q(C)$ are Dirichlet distributions. For simplicity, $D$ and $E$ are treated as fixed parameters. By substituting equations (10) and (11) into equation (4), the variational free energy for the considered POMDPs is obtained as follows:

$$\mathcal{F} = \sum_{\tau=1}^{t} \mathbf{s}_\tau \cdot \{\ln \mathbf{s}_\tau - \ln \mathbf{A} \cdot o_\tau - \ln \mathbf{B} \, \boldsymbol{\psi}_{\tau-1}^s\}$$

$$+ \sum_{\tau=1}^{t} \boldsymbol{\delta}_\tau \cdot \{\ln \boldsymbol{\delta}_\tau - (\vec{1} - 2\Gamma_\tau) \odot \ln \mathbf{C} \, \boldsymbol{\psi}_{\tau-1}^\delta\} + \mathcal{D}_{\mathrm{KL}}[Q(\theta)||P(\theta)] \qquad (12)$$

Here, the Kullback–Leibler divergence $\mathcal{D}_{\mathrm{KL}}[Q(\theta)||P(\theta)]$ represents the complexity of the parameters, which is in the order of $\ln t$ and thus negligibly smaller than the leading order term. When written explicitly, it is given as $\mathcal{D}_{\mathrm{KL}}[Q(\theta)||P(\theta)] = \mathcal{D}_{\mathrm{KL}}[Q(A)||P(A)] + \mathcal{D}_{\mathrm{KL}}[Q(B)||P(B)] + \mathcal{D}_{\mathrm{KL}}[Q(C)||P(C)]$.

Each element of $\boldsymbol{\vartheta}$ is optimised following the gradient descent on $\mathcal{F}$, $\dot{\boldsymbol{\vartheta}} \propto -\partial_{\boldsymbol{\vartheta}}\mathcal{F}$. The posterior expectations about the hidden states and actions—in the form of a Bayesian filter—are derived as the fixed-point solution of implicit gradient descent $\partial_{\mathbf{s}_t}\mathcal{F} = 0$ and $\partial_{\boldsymbol{\delta}_t}\mathcal{F} = 0$, which are given as follows:

$$\begin{cases} \mathbf{s}_t = \sigma(\ln \mathbf{A} \cdot o_t + \ln \mathbf{B}\, \boldsymbol{\psi}^s_{t-1}) \\ \quad \boldsymbol{\delta}_t = \sigma(\ln \mathbf{C}\, \boldsymbol{\psi}^\delta_{t-1}) \end{cases} \quad (13)$$

where $\sigma$ denotes the soft-max function. The updates of $\mathbf{s}_t$ and $\boldsymbol{\delta}_t$ represent state transition and memory readout of the internal Turing machine, respectively. Because $\mathbf{s}_t$ is updated depending on the sensory input term $\ln \mathbf{A} \cdot o_t$, this configuration makes the internal Turing machine's states ($\mathbf{s}_t$) encode the external Turing machine's states ($s_t$). Moreover, these update rules are a family of equation (1) and thus formally associated with the neural activity, as described in the subsequent section.

Moreover, from $\partial_\theta \mathcal{F} = 0$, the posterior expectations about memory and parameters are updated as follows:

$$\begin{cases} \mathbf{a} \leftarrow \mathbf{a} + o_t \otimes \mathbf{s}_t \\ \mathbf{b} \leftarrow \mathbf{b} + \mathbf{s}_t \otimes \boldsymbol{\psi}^s_{t-1} \\ \mathbf{c} \leftarrow \sigma(\mathbf{c}) + (\vec{1} - 2\Gamma_t) \odot \boldsymbol{\delta}_t \otimes \boldsymbol{\psi}^\delta_{t-1} \end{cases} \quad (14)$$

The belief update of $C$ implements a memory writing rule, which is expressed as a risk-modulated policy update rule in the form of modulated Hebbian plasticity [20]. Because the prior term $\sigma(\mathbf{c})$ is restricted within a small value, $C$ has a high learning rate and is thus quickly updated. The update of $\mathbf{c}$ is a simple Hebbian rule when the risk is low; thus, the current memory state is retained. Conversely, it becomes an anti-Hebbian rule when the risk is high, where the memory state is flipped. This enables the writing of new information into the memory. Because the value of writing is determined based on past hidden states and actions, the memory can be viewed as a compressed storage of past state/action information.

**Canonical neural networks**

In this section, the equivalence between canonical neural networks and Turing machines is elaborated by extending the natural equivalence between canonical neural networks and variational Bayesian inference under a class of POMDP generative models [19,20].

Canonical neural networks comprise a two-layer network of rate-coding neurons, where the network's internal states $\{x_{1:t}, y_{1:t}, \omega, \phi\}$ include neural activity $u_t = \{x_t, y_t\}$, synaptic weights $\omega = \{W, K, V\}$, and any other free parameters $\phi = \{\phi^x, \phi^y\}$ that characterise the Helmholtz energy $\mathcal{A}$. Upon receiving sensory inputs $o_t$, the middle-layer ($x_t$) and output-layer ($y_t$) neural activity is given as follows:

$$\begin{cases} \dot{x}_t \propto -\text{sig}^{-1}(x_t) + W o_t + K \psi_{t-1}^x + h^x \\ \dot{y}_t \propto -\text{sig}^{-1}(y_t) + V \psi_{t-1}^y + h^y \end{cases} \tag{15}$$

where $\psi_t^x = \psi(x_t, y_t)$ and $\psi_t^y = \psi(x_t, y_t, \psi_{t-1}^y)$ denote basis functions that summarise the middle and output layer neural activity; $W = W_1 - W_0$, $W = K_1 - K_0$, and $V = V_1 - V_0$ are synaptic weights; and $h^x = h_1^x - h_0^x$ and $h^y = h_1^y - h_0^y$ are the adaptive firing thresholds, which are functions of the synaptic weights. One may consider that $W_1, K_1, V_1$ are excitatory synapses, whereas $W_0, K_0, V_0$ are inhibitory synapses. Both the middle and output layers involve recurrent circuits with one-time-step delays. The output-layer activity $y_t$ determines actions or decisions $\delta_t$. The fixed point of equation (15) provides rate-coding models with a widely used sigmoidal activation function, also known as a neurometric function [59].

By taking the integral of equation (15), a biologically plausible Helmholtz energy for canonical neural networks is obtained [19,20]:

$$\mathcal{A} = \sum_{\tau=1}^{T} \left(\frac{x_\tau}{\overline{x_\tau}}\right)^{\mathrm{T}} \left\{ \ln\left(\frac{x_\tau}{\overline{x_\tau}}\right) - \binom{W_1}{W_0} o_\tau - \binom{K_1}{K_0} \psi_{\tau-1}^x - \binom{h_1^x}{h_0^x} \right\}$$

$$+ \sum_{\tau=1}^{T} \left(\frac{y_\tau}{\overline{y_\tau}}\right)^{\mathrm{T}} \left\{ \ln\left(\frac{y_\tau}{\overline{y_\tau}}\right) - \binom{\vec{1} - 2\Gamma_\tau}{\vec{1} - 2\Gamma_\tau} \odot \binom{V_{\tau 1}}{V_{\tau 0}} \psi_{\tau-1}^y - \binom{h_1^y}{h_0^y} \right\} + \mathcal{C} \quad (16)$$

where $\Gamma_t$ denotes the risk associated with future outcomes. Here, $\Gamma_t$ is considered to modulate Hebbian plasticity without delay due to high plasticity rate. This indicates that canonical neural networks are characterised by Hamiltonian $\mathcal{H}_\xi$ encoded by gene $\xi$ [21]. The expectation of $\mathcal{H}_\xi$ is given as:

$$\langle \mathcal{H}_\xi \rangle = -\sum_{\tau=1}^{t} \left[ \left(\frac{x_\tau}{\overline{x_\tau}}\right)^{\mathrm{T}} \left\{ \binom{W_1}{W_0} o_\tau + \binom{K_1}{K_0} \psi_{\tau-1}^x + \binom{h_1^x}{h_0^x} \right\} + \left(\frac{y_\tau}{\overline{y_\tau}}\right)^{\mathrm{T}} \left\{ \binom{\vec{1} - 2\Gamma_\tau}{\vec{1} - 2\Gamma_\tau} \odot \binom{V_{\tau 1}}{V_{\tau 0}} \psi_{\tau-1}^y + \binom{h_1^y}{h_0^y} \right\} \right] \quad (17)$$

When neural spikes follow Bernoulli distributions, substituting equation (17) into equation (2) yields equation (16). One may associate $e^{-\beta \mathcal{H}_\xi}$ with the probability distribution of spiking neural activity patterns.

Minimising $\mathcal{A}$ furnishes the dynamics of neural networks, including their activity and plasticity. The gradient descent on $\mathcal{A}$ with respect to $x_t$ and $y_t$, i.e., $\dot{x}_t \propto -\partial_{x_t}\mathcal{A}$ and $\dot{y}_t \propto -\partial_{y_t}\mathcal{A}$, derives equation (15) by construction. Moreover, the gradient descent on $\mathcal{A}$ with respect to synaptic weights $\{W_1, W_0, K_1, K_0, V_1, V_0\}$, i.e., $\dot{W}_l \propto -\partial_{W_l}\mathcal{A}$, $\dot{K}_l \propto -\partial_{K_l}\mathcal{A}$, and $\dot{V}_l \propto -\partial_{V_l}\mathcal{A}$ (for $l = 1, 0$), provides the following synaptic plasticity rules:

$$\begin{cases} \dot{W}_1 \propto \langle x_t o_t^{\mathrm{T}} \rangle - \langle x_t \vec{1}^{\mathrm{T}} \rangle \odot \mathrm{sig}(W_1) \\ \dot{W}_0 \propto \langle \overline{x_t} o_t^{\mathrm{T}} \rangle - \langle \overline{x_t} \vec{1}^{\mathrm{T}} \rangle \odot \mathrm{sig}(W_0) \\ \dot{K}_1 \propto \langle x_t {\psi_{t-1}^x}^{\mathrm{T}} \rangle - \langle x_t \vec{1}^{\mathrm{T}} \rangle \odot \mathrm{sig}(K_1) \\ \dot{K}_0 \propto \langle \overline{x_t} {\psi_{t-1}^x}^{\mathrm{T}} \rangle - \langle \overline{x_t} \vec{1}^{\mathrm{T}} \rangle \odot \mathrm{sig}(K_0) \\ \dot{V}_1 \propto \left\langle (\vec{1} - 2\Gamma_t) \odot y_t {\psi_{t-1}^y}^{\mathrm{T}} \right\rangle - \langle y_t \vec{1}^{\mathrm{T}} \rangle \odot \mathrm{sig}(V_1) \\ \dot{V}_0 \propto \left\langle (\vec{1} - 2\Gamma_t) \odot \overline{y_t} {\psi_{t-1}^y}^{\mathrm{T}} \right\rangle - \langle \overline{y_t} \vec{1}^{\mathrm{T}} \rangle \odot \mathrm{sig}(V_0) \end{cases} \quad (18)$$

The three-factor Hebbian plasticity in *V* enables updates of (mental) actions to minimise the risks associated with future outcomes. Neuronal substrates such as neuromodulators [31–33] may encode the risk. The modulated plasticity ensures that only good strategies are accepted, and bad strategies are rejected. As shown in the Results section, this plasticity efficiently rewrites binary memory.

Previous works established that the dynamics of canonical neural networks that minimise a cost function are cast as variational free energy minimisation under a class of factorial POMDP models [19,20], by showing a natural equivalence between the Helmholtz energy $\mathcal{A}$ and variational free energy $\mathcal{F}$. In other words, there exists a generative model that satisfies $\mathcal{A} \equiv \mathcal{F}$, where $\pi(\varphi)$ encodes or parameterises the posterior belief $Q(\vartheta)$, i.e., $\pi(\varphi) \equiv Q(\vartheta)$, and $\xi$ encodes the generative model structure $m = m(\xi)$. As shown previously [20], equation (16) can be transformed into the form in **Fig. 1b**, in which the Hamiltonian formally corresponds to the negative logarithm of the generative model, $\mathcal{H}_\xi = -\ln P_m(o_{1:t}, \vartheta)$. This indicates that any neural network that minimises $\mathcal{A}$ can be conceptualised as performing a variational Bayesian inference. Neural network states $\{x_{1:t}, y_{1:t}, \omega, \phi\}$ correspond to quantities in variational Bayesian inference, such as $x_t \equiv \mathbf{s}_t$, $y_t \equiv \boldsymbol{\delta}_t$, $\omega \equiv \boldsymbol{\theta}$, and $\phi \equiv \lambda$ (**Table 1**). In particular, the memory (policy) matrix **C** corresponds to a synaptic matrix $V$ to the output layer, and $y_t$ represents the memory readout at position $\psi^y_{t-1}$. Owing to this equivalence, internal states of canonical neural networks encode the posterior belief about states of external Turing machines.

However, it should be noted that arbitrarily selected neural dynamics correspond to inferences of the external milieu under a suboptimal generative model with biased prior beliefs. Thus, the generative model must be optimised by updating the genes.

**Evolution as active Bayesian model selection**

This section elaborates on the notion that the Helmholtz free energy minimisation derives natural selection and Bayesian model selection. To characterise the optimal gene distribution in terms of phenotypes, it is necessary to commit to a specific environmental generative process and evolutionary rule. Natural selection updates the current (i.e., prior) distribution $n(\xi)$ to the offspring (i.e., posterior) distribution $N(o_{1:T}, \xi)$, where $T$ denotes the maximum lifetime for an individual. The external milieu states do not directly affect the agent's internal states and vice versa—refer to Markov blanket [60,61]—and the behaviour of each individual affects other individuals only through the environment. Thus, the reproduction rate $\rho(o_{1:T}, \xi)$ is a function of the sensory input sequence $o_{1:T}$ and genes $\xi$.

The maximisation of the offspring number $Z = \sum_{o_{1:T}, \xi} \rho(o_{1:T}, \xi) P(o_{1:T}|\xi) n(\xi)$ can be derived from the Helmholtz energy minimisation at the species or population level. Each agent minimises $\mathcal{A}[\pi(\varphi), o_{1:T}, \xi]$ with respect to the internal state distribution $\pi(\varphi)$. This process can be extended to the population or species level. When doing so, there is a degree of freedom to add an arbitrary function of $o_{1:T}$ and $\xi$, denoted as $\overline{\mathcal{C}}(o_{1:T}, \xi)$, to the Helmholtz energy as the integration constant, because the variational solution of $\pi(\varphi)$ is unchanged by this modification. Thus, the ensemble Helmholtz energy generally has the following form:

$$\overline{\mathcal{A}}[\pi(o_{1:T}, \xi)] = \langle \mathcal{A}[\pi(\varphi), o_{1:T}, \xi] + \overline{\mathcal{C}}(o_{1:T}, \xi) + \ln \pi(o_{1:T}, \xi) \rangle_{\pi(o_{1:T}, \xi)} \quad (19)$$

where $\langle \cdot \rangle_{\pi(o_{1:T}, \xi)} = \sum_{o_{1:T}, \xi} \cdot \pi(o_{1:T}, \xi)$ is the expectation over $\pi(o_{1:T}, \xi)$. The integration constant $\overline{\mathcal{C}}(o_{1:T}, \xi)$ can be read as an arbitrary evolution rule. Because the integration constant $\mathcal{C}$ involved in equation (3) is selected to satisfy $\int e^{-\mathcal{A}} do_{1:t} = 1$ at the steady state, $\mathcal{A}[\pi(\varphi), o_{1:t}, \xi] \geq -\ln P_m(o_{1:t})$ holds. As optimising the internal states $\pi(\varphi)$ provides $\mathcal{A}[\pi(\varphi), o_{1:t}, \xi] \simeq -\ln P_m(o_{1:t})$, solving the variation $\delta \overline{\mathcal{A}} = 0$ with respect to $\pi(o_{1:T}, \xi)$ results in $\pi(o_{1:T}, \xi) \propto P_m(o_{1:T}) e^{-\mathcal{C}(o_{1:T}, \xi)}$.

When genes encode generative models with sufficient representation capacity, $P_m(o_{1:t})$ provides a good approximation of $P(o_{1:t}|m)$ Moreover, $P(o_{1:t}|m) \equiv P(o_{1:t}|\xi)$ holds by construction because model structure $m = m(\xi)$ is a function of $\xi$. More precisely, as $\mathcal{H}_\xi$ can take various form depending on gene $\xi$, there exists a subset of genes that ensure the matching between $P_m(o_{1:t})$ and $P(o_{1:t}|\xi)$. Although initial genes may not be involved in this subset, $\langle -\ln P_m(o_{1:t}) \rangle_{P(o_{1:t},\xi)} \geq \langle -\ln P(o_{1:t}|\xi) \rangle_{P(o_{1:t},\xi)}$ is satisfied from the nonnegativity of the Kullback–Leibler divergence. Thus, the ensemble Helmholtz energy minimisation facilitates the selection of genes that satisfy $P_m(o_{1:t}) \equiv P(o_{1:t}|\xi)$.

Hence, when $\overline{C}(o_{1:T}, \xi) \equiv -\ln \rho(o_{1:T}, \xi) n(\xi)$ is selected, $\pi(o_{1:T}, \xi)$ corresponds to equation (6). Because $\overline{\mathcal{A}} \simeq -\ln Z$ holds with the optimal $\pi(o_{1:T}, \xi)$, this facilitates the most efficient reproduction. The total offspring number $Z$ is maximised by minimising $\mathcal{A}$ owing to equation (9). Therefore, natural selection can be derived from the Helmholtz energy minimisation.

The variational free energy corresponding to the ensemble Helmholtz energy $\overline{\mathcal{A}}$ is given as

$$\overline{\mathcal{F}}[Q(o_{1:T}, m)] = \langle \mathcal{F}[Q(\vartheta), o_{1:T}, m] - \ln \rho(o_{1:T}, m) - \ln n(m) + \ln Q(o_{1:T}, m) \rangle_{Q(o_{1:T},m)} \quad (20)$$

where $\langle \cdot \rangle_{Q(o_{1:T},\xi)}$ indicates the expectation over $Q(o_{1:T}, \xi)$. Solving the variation $\delta \overline{\mathcal{F}} = 0$ provides the approximate posterior distribution $Q(o_{1:T}, m) \propto P_m(o_{1:T}) \rho(o_{1:T}, m) n(m)$. The optimal gene distribution that minimises $\overline{\mathcal{F}}$ guarantees the matching between the generative model and true generative process, $P_m(o_{1:t}, \vartheta) \equiv P(o_{1:t}, \vartheta)$, referred to as Hamiltonian matching [21]. This provides the optimal decision making for survival and reproduction, which makes their reproduction rate superior to that of other genes. Therefore, natural selection can be viewed as an active Bayesian model selection. When the gene mutation probability is sufficiently small, sharp peaks appear at the optimal genes that maximise the offspring number, in which multiple peaks represent equally good solutions.

**Coevolution**

The aforementioned framework can be extended to the coevolution of multiple agents. For simplicity, a case in which *N* agents receive shared sensory inputs $o_t = \{o_t^{(env)}, \delta_t^{(1)}, \dots, \delta_t^{(N)}\}$ is considered, where $\delta_t^{(i)}$ denotes the action or output of the *i*-th agent and $o_t^{(env)}$ denotes signals generated by the environment.

In each generation, each agent updates the input and gene distribution from $P(o_{1:T}, m)$ to $Q(o_{1:T}, m) \propto P(o_{1:T}, m)\rho_i(o_{1:T}, m)^\varepsilon$, where $\varepsilon$ is a small positive constant and $P(o_{1:T}, m)$ can be read as a prior distribution. The reproduction rate $\rho_i(o_{1:t}, m)$ of each agent can be decomposed into cooperative $\rho^*(o_{1:t}, m)$ and competitive $\Delta\rho_i(o_{1:t}, m)$ components, which satisfy $\rho_i = \rho^* + \Delta\rho_i$ and $\sum_{i=1}^{N} \Delta\rho_i = 0$. The competitive component corresponding to a zero-sum game. In this setup, for one round of natural selection, the posterior (offspring) distribution is expressed as

$$Q(o_{1:T}, m) \propto P(o_{1:T}|m)P(m) \prod_{i=1}^{N}(\rho^* + \Delta\rho_i)^\varepsilon \qquad (21)$$

The limit of this evolutionary process can be analysed as follows: from the AM-GM inequality, $\rho^*(o_{1:T}, \xi)^N \geq \prod_{i=1}^{N} \rho_i(o_{1:T}, \xi)$ holds. This indicates that only the agent with the cooperative term (i.e., $\xi$ with $\Delta\rho_i = 0$) becomes the primary trend, whereas the competitive term hinders maximising the reproduction. Thus, when the gene mutation probability is sufficiently small, equation (21) asymptotically converges to a steady-state distribution with sharp peaks, $Q(o_{1:T}, m) \propto \lim_{n \to \infty} \rho^*(o_{1:T}, m)^n$, which is characterised only by the cooperative component. These observations imply that survival agents select strategies that are beneficial for the survival of the species, but not for the individual. In other words, it converges to a solution that optimises the non-zero-sum game.

**Numerical simulations**

In **Fig. 2**, canonical neural networks inferred states of external adders. Based on handwritten digit inputs, the networks inferred the hidden states ($s_t$) and memory ($C$) and learned state transition mapping ($B$). For simplicity, the likelihood mapping ($A$) was given a priori. The update rules were given as the gradient descent on $\mathcal{A}$, as shown in equation (15), which include the state update of the automaton and memory readout. Reading header movement was expressed by basis functions $\psi_t^\delta$. For simplicity, random-access memory that can freely move header position, expressed as $\psi_t^\delta = H^\delta s_t$, was adopted in this simulation. Memory writing occurred through the risk modulated Hebbian plasticity of $V$ following equation (18), where risk $\Gamma_t = \sigma\left(\left(\psi_{t1}^\gamma, \psi_{t2}^\gamma + \psi_{t3}^\gamma + \psi_{t4}^\gamma\right)^\mathrm{T}\right)$ was characterised by bases $\psi_t^\gamma = \left(s_{t,1:16}^\mathrm{T} s_{t,17:32}, s_{t-1,1:16}^\mathrm{T} s_{t-1,17:32}, \delta_{t-1}, \Gamma_{t-1}\right)^\mathrm{T}$.

In **Fig. 3**, the reproduction rate varied depending on the performance of estimating the summation of input binary numbers. Agents employed a genetically encoded risk $\Gamma_t$ defined as a function of bases, $\Gamma_t = \left(\xi_1\psi_{t1}^\gamma + \xi_3\psi_{t2}^\gamma + \xi_5\psi_{t3}^\gamma + \xi_7\psi_{t4}^\gamma, \xi_2\psi_{t1}^\gamma + \xi_4\psi_{t2}^\gamma + \xi_6\psi_{t3}^\gamma + \xi_8\psi_{t4}^\gamma,\right)^\mathrm{T}$, which was characterised by an eight-bit gene vector $\xi \in \{0,1\}^8$. When the corresponding gene adopted the value of 1, the term was included in the risk. When a gene bears $\xi = (1,0,0,1,0,1,0,1)^\mathrm{T}$, the risk suits to act as an adder and thus maximise the reproduction rate. The gene distribution was initialised with a uniform distribution. Natural selection was depicted as an operation that adds small mutation to the parent gene distribution $n(\xi)$ to create $n'(\xi)$ and then replaces $n'(\xi)$ with the offspring gene distribution $\pi(o_{1:t}, \xi)$ following equation (6). This process was iterated for 40 generations.

In **Fig. 4**, the environment was determined by 10 Turing machines, each of which involved 10-dimensional hidden states ($s_t$) and two-dimensional mental action ($a_t$) in the form of a POMDP, in which $s_t$ and $a_t$ are one-hot vectors. The transition matrix $B \in \{0,1\}^{10 \times 10 \times 2}$ was randomly generated for each Turing machine, whereas the likelihood mapping $A \in [0,1]^{10 \times 10}$ was randomly generated and commonly used for all Turing machines. The canonical neural network employed two actions $y_t$ and $y'_t$, where $y_t$ encoded memory readout from $V$, while $y'_t$ encoded readout from $V'$ that memorises transition mappings of external Turing machines.

**Data Availability**

All relevant data are presented in this paper. Figs. 2–4 were generated using the author's scripts (see Code Availability).

**Code Availability**

The MATLAB scripts are available at https://github.com/takuyaisomura/reverse_engineering. The scripts are covered under the GNU General Public Licence v3.0.

**References**

1. Turing, A. M. On computable numbers, with an application to the Entscheidungsproblem. *Proc. London Mathematical Society* **S2–42**, 230–265 (1936).

2. Hodgkin, A. L. & Huxley, A. F. A quantitative description of membrane current and its application to conduction and excitation in nerve. *J. Physiol.* **117**, 500–544 (1952).

3. FitzHugh, R. Impulses and physiological states in theoretical models of nerve membrane. *Biophys. J.* **1**, 445–466 (1961).

4.  Nagumo, J., Arimoto, S. & Yoshizawa, S. An active pulse transmission line simulating nerve axon. *Proc. IRE* **50**, 2061–2070 (1962).

5.  Hebb, D. O. *The Organization of Behavior: A Neuropsychological Theory* (Wiley, New York, 1949).

6.  Song, S., Miller, K. D. & Abbott, L. F. Competitive Hebbian learning through spike-timing-dependent synaptic plasticity. *Nat. Neurosci.* **3**, 919–926 (2000).

7.  Clopath, C., Büsing, L., Vasilaki, E. & Gerstner, W. Connectivity reflects coding: a model of voltage-based STDP with homeostasis. *Nat. Neurosci.* **13**, 344–352 (2010).

8.  Roth, G. & Dicke, U. Evolution of the brain and intelligence. *Trends Cogn. Sci*. **9**, 250–257 (2005).

9.  Rabinovich, M. I., Varona, P., Selverston, A. I. & Abarbanel, H. D. Dynamical principles in neuroscience. *Rev. Mod. Phys.* **78**, 1213–1265 (2006).

10. Sussillo, D. & Abbott, L. F. Generating coherent patterns of activity from chaotic neural networks. *Neuron* **63**, 544–557 (2009).

11. Laje, R. & Buonomano, D. V. Robust timing and motor patterns by taming chaos in recurrent neural networks. *Nat. Neurosci.* **16**, 925–933 (2013).

12. Knill, D.C. & Pouget, A. The Bayesian brain: the role of uncertainty in neural coding and computation. *Trends Neurosci.* **27**, 712–719 (2004).

13. Doya, K., Ishii, S., Pouget, A. & Rao, R. P. (Eds.) *Bayesian Brain: Probabilistic Approaches to Neural Coding* (MIT Press, Cambridge, MA, USA, 2007).

14. Linsker, R. Self-organization in a perceptual network. *Computer* **21**, 105–117 (1988).

15. Dayan, P., Hinton, G. E., Neal, R. M. & Zemel, R. S. The Helmholtz machine. *Neural Comput.* **7**, 889–904 (1995).

16. Sutton, R. S. & Barto, A. G. *Reinforcement learning* (Cambridge, MA: MIT Press, 1998).

17. Friston, K. J., Kilner, J. & Harrison, L. A free energy principle for the brain. *J. Physiol. Paris* **100**, 70–87 (2006).

18. Friston, K. J. The free-energy principle: a unified brain theory? *Nat. Rev. Neurosci.* **11**, 127–138 (2010).

19. Isomura, T. & Friston, K. J. Reverse-engineering neural networks to characterize their cost functions. *Neural Comput.* **32**, 2085–2121 (2020).

20. Isomura, T., Shimazaki, H. & Friston, K. J. Canonical neural networks perform active inference. *Commun. Biol.* **5**, 55 (2022).

21. Isomura, T. Bayesian mechanics of self-organising systems. arXiv:2311.10216 (2023).

22. Tazawa, U. T. & Isomura, T. Synaptic pruning facilitates online Bayesian model selection. Preprint at *bioRxiv* https://www.biorxiv.org/content/10.1101/2024.05.15.593712v1 (2024).

23. Isomura, T., Kotani, K. & Jimbo, Y. Cultured cortical neurons can perform blind source separation according to the free-energy principle. *PLoS Comput. Biol.* **11**, e1004643 (2015).

24. Isomura, T. & Friston, K. J. In vitro neural networks minimise variational free energy. *Sci. Rep.* **8**, 16926 (2018).

25. Isomura, T., Kotani, K., Jimbo, Y. & Friston, K. J. Experimental validation of the free-energy principle with in vitro neural networks. *Nat. Commun.* **14**, 4547 (2023).

26. Graves, A. Neural Turing Machines. Preprint at *arXiv* arXiv:1410.5401. https://arxiv.org/abs/1410.5401 (2014).

27. Graves, A., Wayne, G., Reynolds, M. et al. Hybrid computing using a neural network with dynamic external memory. *Nature* **538**, 471–476 (2016).

28. Cabessa, J. Turing complete neural computation based on synaptic plasticity. *PLoS One* **14**, e0223451 (2019).

29. Cabessa, J. & Tchaptchet, A. Automata complete computation with Hodgkin–Huxley neural networks composed of synfire rings. *Neural Netw.* **126**, 312–334 (2020).

30. Plank, J., Zheng, C., Schuman, C. & Dean, C. Spiking neuromorphic networks for binary tasks. International Conference on Neuromorphic Systems **22**, 1–9 (2021).

31. Pawlak, V., Wickens, J. R., Kirkwood, A. & Kerr, J. N. Timing is not everything: neuromodulation opens the STDP gate. *Front. Syn. Neurosci.* **2**, 146 (2010).

32. Frémaux, N. & Gerstner, W. Neuromodulated spike-timing-dependent plasticity, and theory of three-factor learning rules. *Front. Neural Circuits* **9**, 85 (2016).

33. Kuśmierz, Ł., Isomura, T. & Toyoizumi, T. Learning with three factors: modulating Hebbian plasticity with errors. *Curr. Opin. Neurobiol.* **46**, 170–177 (2017).

34. Wald, A. An essentially complete class of admissible decision functions. *Ann. Math. Stat.* **18**, 549–555 (1947).

35. Brown, L. D. A complete class theorem for statistical problems with finite-sample spaces. *Ann. Stat.* **9**, 1289–1300 (1981).

36. Berger, J. O. *Statistical Decision Theory and Bayesian Analysis* (Springer Science & Business Media, Berlin, 2013).

37. Blei, D. M., Kucukelbir, A. & McAuliffe, J. D. Variational inference: A review for statisticians. *J. Am. Stat. Assoc.* **112**, 859–877 (2017).

38. Friston, K. J., FitzGerald, T., Rigoli, F., Schwartenbeck, P. & Pezzulo, G. Active inference: a process theory. *Neural Comput.* **29**, 1–49 (2017).

39. Sajid, N., Ball, P. J., Parr, T. & Friston, K. J. Active inference: demystified and compared. *Neural Comput.* **33**, 674–712 (2021).

40. Parr, T., Pezzulo, G. & Friston, K. J. *Active inference: the free energy principle in mind, brain, and behavior.* (MIT Press, Cambridge, MA, USA, 2022).

41. Isomura, T., Parr, T. & Friston, K. J. Bayesian filtering with multiple internal models: toward a theory of social intelligence. *Neural Comput.* **31**, 2390–2431 (2019).

42. Kobayashi, T. J. & Sughiyama, Y. Fluctuation relations of fitness and information in population dynamics. *Phys. Rev. Lett.* **115**, 238102 (2015).

43. Kobayashi, T. J. & Sughiyama, Y. Stochastic and information-thermodynamic structures of population dynamics in a fluctuating environment. *Phys. Rev. E* **96**, 012402 (2017).

44. Nakashima, S. & Kobayashi, T. J. Acceleration of evolutionary processes by learning and extended Fisher's fundamental theorem. *Phys. Rev. Res.* **4**, 013069 (2022).

45. Shimazaki, H. & Niebur, E. Phase transitions in multiplicative competitive processes. *Phys. Rev. E* **72**, 011912 (2005).

46. Kirchhoff, M., Parr, T., Palacios, E., Friston, K. J. & Kiverstein, J. The Markov blankets of life: autonomy, active inference and the free energy principle. *J. R. Soc. Interface* **15**, 20170792 (2018).

47. Constant, A., Ramstead, M. J. D., Veissiere, S. P., Campbell, J. O. & Friston, K. J. A variational approach to niche construction. *J. R. Soc. Interface* **15**, 20170685 (2018).

48. Ramstead, M. J. D., Constant, A., Badcock, P. B. & Friston, K. J. Variational ecology and the physics of sentient systems. *Phys. Life Rev.* **31**, 188–205 (2019).

49. Pezzulo, G., Parr, T. & Friston, K. J. The evolution of brain architectures for predictive coding and active inference. *Philos. Trans. R. Soc. B* **377**, 20200531 (2022).

50. Rogozhin, Y. Small universal Turing machines. *Theoret. Comput. Sci.* **168**, 215–240 (1996).

51. Neary, T. & Woods, D. Four small universal Turing machines. Proc. 5th Int. Conf. on Machines, Computations, and Universality, LNCS-4664, Springer-Verlag, 242–254 (2007).

52. Mirza, M. B., Adams, R. A., Friston, K. J. & Parr, T. Introducing a Bayesian model of selective attention based on active inference. *Sci. Rep.* **9**, 13915 (2019).

53. Gershman, S. J., Assad, J. A., Datta, S. R. et al. Explaining dopamine through prediction errors and beyond. *Nat. Neurosci.* **27**, 1645–1655 (2024).

54. Neves, G., Cooke, S. F. & Bliss, T. V. Synaptic plasticity, memory and the hippocampus: a neural network approach to causality. *Nat. Rev. Neurosci.* **9**, 65–75 (2008).

55. Isomura, T. Quadratic speedup of global search using a biased crossover of two good solutions. Preprint at *arXiv* arXiv:2111.07680. https://arxiv.org/abs/2111.07680 (2021).

56. Friston, K. & Frith, C. A duet for one. *Conscious. Cogn.* **36**, 390–405 (2015).

57. Friston, K. J., Parr, T., Heins, C. et al. Federated inference and belief sharing. *Neurosci. Biobehav. Rev.* **156**, 105500 (2023).

58. Friston, K. J., Stephan, K. E., Montague, R. & Dolan, R. J. Computational psychiatry: the brain as a phantastic organ. *Lancet Psychiatry* **1**, 148–158 (2014).

59. Newsome, W. T., Britten, K. H. & Movshon, J. A. Neuronal correlates of a perceptual decision. *Nature* **341**, 52–54 (1989).

60. Da Costa, L., Parr, T., Sajid, N., Veselic, S., Neacsu, V. & Friston, K. J. Active inference on discrete state-spaces: A synthesis. *J. Math. Psychol.* **99**, 102447 (2020).

61. Friston, K. J., Da Costa, L., Sajid, N. et al. The free energy principle made simpler but not too simple. *Phys. Rep.* **1024**, 1–29 (2023).

**Acknowledgements**

T.I. is supported by the Japan Society for the Promotion of Science (JSPS) KAKENHI Grant Number JP23H04973 and the Japan Science and Technology Agency (JST) CREST Grant Number JPMJCR22P1. The funders had no role in study design, data collection and analysis, decision to publish, or preparation of the manuscript.

**Competing interest declaration**

The author has no competing interests to declare.

**Table 1. Correspondence of components of canonical neural networks, Bayesian inference, and turning machines**

| Neural network formation | Turing machine formation | Bayesian inference (POMDP) formation |
|---|---|---|
| Middle-layer neural activity $x_t$ | Automaton states $x_t$ | Hidden state posterior $\mathbf{s}_t$ |
| Output-layer neural activity $y_t$ | Memory readout $y_t$ | Action posterior $\boldsymbol{\delta}_t$ |
| Neuromodulator $\varGamma_t$ | Memory writing signal $\varGamma_t$ | Risk $\varGamma_t$ |
| Basis function $\psi_t^x$ | Basis function $\psi_t^x$ | Basis function $\boldsymbol{\psi}_t^s$ |
| Basis function $\psi_t^y$ | Reading header position $\psi_t^y$ | Basis function $\boldsymbol{\psi}_t^\delta$ |
| Synaptic weights in middle layer $W_1, W_0$ | Signal weights from sensory inputs $W_1, W_0$ | Likelihood mapping posterior $\mathbf{A}$ |
| Synaptic weights in middle layer $K_1, K_0$ | Transition mapping $K_1, K_0$ | Transition mapping posterior $\mathbf{B}$ |
| Synaptic weights in output layer $V_1, V_0$ | Memory matrix $V_1, V_0$ | Policy mapping posterior $\mathbf{C}$ |
| Gene $\xi$ | Program $\xi$ | Model structure $m$ |
| Internal state distribution $\pi$ | Internal state distribution $\pi$ | Posterior distribution $Q$ |